\documentclass[12pt]{article}

\usepackage{epsfig}

\begin{document}

\begin{flushright}
  CPHT-RR 007.0304
\end{flushright}

\vspace{\baselineskip}

\begin{center}
\textbf{\Large
Power corrections to the process $\gamma^{*}\gamma \to \pi\pi$ in the
running coupling method} \\

\vspace{2\baselineskip}

{\large S. S. Agaev$^{a,b}$,\, M. Guidal$^b$ , B.~Pire$^c$} \\

\vspace{3\baselineskip}

${}^a$\,High Energy Physics Lab.,
               Baku State University, 370148 Baku, Azerbaijan
\\[0.5\baselineskip]
${}^b$\,Institut de Physique Nucl\'{e}aire Orsay, F-91406 Orsay, France
\\[0.5\baselineskip]
${}^c$\,CPHT\footnote{UnitŽ mixte de recherche du CNRS (UMR 7644)}, {\'E}cole Polytechnique, 91128 Palaiseau, France
\\[0.5\baselineskip]

\vspace{4\baselineskip}

\textbf{Abstract}\\

\vspace{1\baselineskip}

\parbox{0.9\textwidth} {Power-suppressed corrections  coming from the
end-point integration regions in the amplitude of the process
$\gamma^{*}\gamma \to \pi \pi$ at large $Q^2$ and small squared center-of-mass energy
$W^2$ are calculated in the QCD hard-scattering approach where the amplitudes factorize
in a hard perturbatively calculable part and a generalized distribution amplitude.
The running coupling method and the technique of  infrared
renormalon calculus are applied to obtain Borel resummed expressions
for the two main components of the process amplitude.  Numerical estimates
for these power corrections are presented. They are sizeable when $Q^2 < 10 \;GeV^2$.}
\end{center}

\vspace{1\baselineskip}

\section{Introduction}

Single meson and meson pair productions with a small invariant
mass $W$ in virtual photon-photon collisions are  exclusive
processes for which perturbative QCD (PQCD) \cite{BL80} analysis
was successfully applied when the virtuality $Q^2$ of one photon
is high. These investigations are based on the perturbative QCD
factorization theorems which allow one to compute the amplitude of
the exclusive process as the convolution integral of the meson
distribution amplitude (DA) or the two-meson generalized
distribution amplitude (GDA) \cite{P98} and the hard-scattering
amplitude of the underlying partonic subprocess. The meson DA's
and GDA's $\phi(x,\mu_F^2)$ and $\phi(z,\zeta,W^2, \mu_F^2)$ are
non-perturbative objects and contain the long-distance mesonic
binding and hadronization effects. The GDA's are related by
crossing \cite{Teryaev:2001qm} to generalized parton distributions
\cite{JiRad,DiehlPRep}. The theoretical and experimental
investigations of the processes $\gamma^{*}\gamma \to
M\overline{M}$ open opportunities to obtain new, valuable
information on the fragmentation of quarks or gluons into mesons
\cite{PS}. These processes can be studied in $e\gamma$ and
$e^{+}e^{-}$ collisions and first results have been published
\cite{exp}.

The perturbative QCD approach  and the factorization theorems
describe  exclusive processes at asymptotically large values of
the squared momentum transfer $Q^2$. But in the present
experimentally accessible energy regimes, power-suppressed
corrections may play an important role. There are numerous sources
of power corrections to the  process $\gamma^{*}\gamma\to
M\overline{M}$. For example, power corrections arise due to the
intrinsic transverse momentum of partons retained in the
corresponding subprocess hard-scattering amplitude and GDA's. They
may be estimated along the lines presented in Ref.\ \cite{G99},
where such corrections were calculated for the deeply virtual
electroproduction of photon and mesons on the nucleon. A power
correction to the process $\gamma^{*}\gamma \to \pi \pi$ has been
estimated for the  amplitude corresponding to scattering of two
photons with equal helicities, with the help of the light-cone sum
rules method \cite{Kiv00}.
 The power-suppressed (twist-3) contribution due to the interaction of a
longitudinally polarized virtual photon with the real one
was analyzed within the Wandzura-Wilczek approximation in Ref.\ \cite{Kiv01}.

In the present paper we compute a class of power corrections which originate
from the end-point regions $z\rightarrow 0,\,1$ in the integration of the PQCD
factorization expression over the parton
longitudinal momentum fraction $z$. We restrict ourselves to  the two-pion final state and
to the leading twist-2 amplitudes.
Generalization of our approach to encompass other two-meson final states is
straightforward.

It has been advocated \cite{BL83} that, in order to reduce the higher-order
corrections to a physical quantity and improve the convergence of the
corresponding perturbation series, the renormalization scale, i.e.
the argument of the QCD coupling in a Feynman
diagram  should be set equal to the virtual parton's
squared four-momentum. In exclusive
processes, the scale
chosen this way inevitably depends on the longitudinal momentum
fractions carried by the hadron constituents. In our case,
the relevant scale is given by $\mu _R^2=Q^2z$ or
$Q^2\overline{z}$ \cite{P98}. But then the PQCD factorization formula
diverges, since $\alpha _{{\rm s}}(Q^2z)$ [$\alpha _{{\rm s}}(Q^2\overline{z})$]
suffers from an end-point $z\to  0$ [$z \to 1$] singularity. This problem may be
solved by freezing the argument of the QCD coupling and performing all calculations with
$\alpha _{{\rm s}}(Q^2)$ [or $\alpha _{{\rm s}}(Q^2/2)$]. In the running coupling (RC)
method, one allows the argument of $\alpha _{{\rm s}}$ to run but removes divergences
appearing in the perturbative expression with the help of a Borel transformation and a
principal value prescription. It turns out that this procedure, used in conjunction with
the infrared (IR) renormalon technique \cite{ben,gr} allows one to obtain the Borel
resummed expression for the process amplitude and  estimate power corrections arising
from the end-point integration regions. This method was used for other processes
\cite{ag4,ag1,ag3,ag2,ag6} and succesfully confronted to experimental data.

This paper is organized as follows: in Sect. 2 we present
kinematics, general expressions for the amplitude of the process
and the two-pion GDA's. In Sect. 3 we outline the main points of
the RC method and obtain the Borel resummed components of the
amplitude. Section 4 contains results of our numerical
calculations. Finally, we give our conclusions in Sect. 5.

\section{Amplitude of the process  $\gamma^{*}\gamma \to \pi \pi$}

\begin{figure}
\epsfxsize=10 cm
\epsfysize=8 cm
\centerline{\epsffile{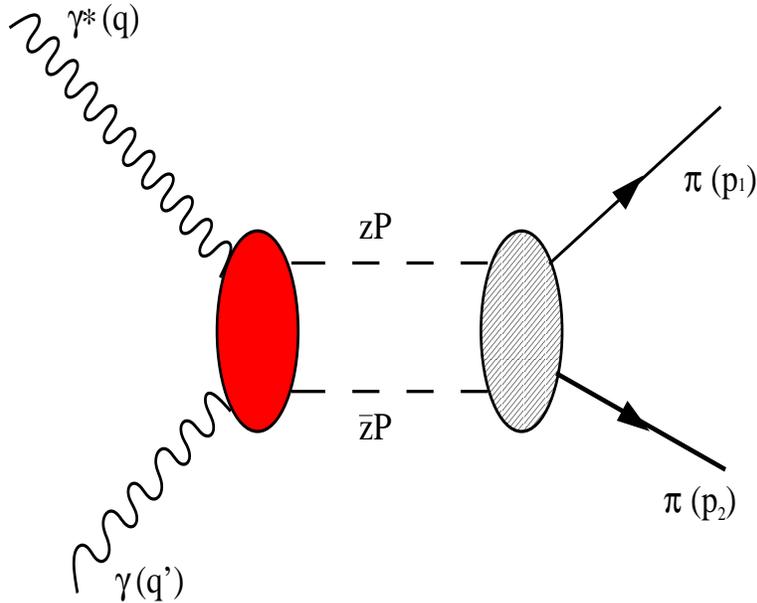}}
\caption{\label{fig:pros} Schematical representation of the
factorization theorem for the process $\gamma^{*}\gamma \to \pi \pi$.
The solid and dashed blobs denote the hard-scattering subprocess
$\gamma^{*}\gamma \to q\overline{q}$ ($\gamma^{*}\gamma \to gg$) and
hadronization $q\overline{q} \to \pi \pi$
($gg \to \pi \pi$), respectively. The quarks (gluons) are depicted as dashed lines.
The momenta of the initial and final particles are shown in the figure. The
total momentum of the final state is $P=p_1+p_2$ and
$z$ is the longitudinal momentum fraction of the $2\pi$
system carried by the quark (gluon).}
\end{figure}

The process

\begin{equation}
\label{eq:1}\gamma^{*}(q)+\gamma (q^{\prime}) \to \pi(p_1) +\pi(p_2),
\end{equation}
is schematically depicted in Fig.\ \ref{fig:pros}. In addition to the four-momenta of the initial and
final particles  one introduces the total and relative momenta of the pion pair
$P=p_1+p_2,\;\;\Delta=p_2-p_1$.
The momenta of the involved particles can be described in
terms of two light-like vectors $p,\,n$ which obey $p \cdot n=1$.
The decomposition of the four-momenta of the initial and final
states in terms of the vectors $p$ and $n$ reads

$$
q=p- \frac{Q^2}{2}n,\;\; q^{\prime}= \frac{Q^2+W^2}{2}n, \; \; q^2=-Q^2,\,\;q^{\prime 2}=0,
$$

$$
p_1=\zeta p +\overline{\zeta}\frac{W^2}{2}n-\frac{k_{\perp}}{2},\;\;
p_2=\overline{\zeta}p+ \zeta\frac{W^2}{2}n+\frac{k_{\perp}}{2},
$$

\begin{equation}
\label{eq:2}
P=p+\frac{W^2}{2}n,\;P^2=W^2,\;
\Delta=(\overline{\zeta}-\zeta)p+(\zeta-\overline{\zeta})\frac{W^2}{2}n+k_{\perp},
\end{equation}
where the quantities

$$
\zeta=\frac{p_1\cdot n}{P\cdot n},\;\;\overline{\zeta}=1-\zeta = \frac{p_2\cdot
n}{P\cdot n},
$$
describe the distribution of the longitudinal momentum between two
pions. The vectors $p$ and $n$ can also be employed to define the
metric tensor in the transverse space:

\begin{equation}
\label{eq:3}
(-g^{\mu \nu})_{T}=-g^{\mu \nu}+ p^{\mu}n^{ \nu}+
p^{\nu}n^{ \mu}.
\end{equation}
In the hard photoproduction regime $Q^2 \gg W^2 , \Lambda^2$ the
amplitude of the process (\ref{eq:1}) has the form

\begin{equation}
\label{eq:4}
T^{\mu \nu}(\zeta,W^2)=\frac{i}{2}\left(-g^{\mu \nu} \right)_TT_0(\zeta,W^2)+
\frac{i}{2}\frac{k_{\perp}^{\nu}(P+q^{\prime})^{\mu}}{Q^2}T_1(\zeta,W^2)+
\frac{i}{2}\frac{k_{\perp}^{( \mu}k_{\perp}^{\nu )}}{W^2}T_2(\zeta,W^2),
\end{equation}
where $k_{\perp}^{( \mu}k_{\perp}^{\nu )}$ is the traceless, symmetric
tensor product of the relative transverse  momentum of the pion pair

$$
k_{\perp}^{( \mu}k_{\perp}^{\nu )}=k_{\perp}^{ \mu}k_{\perp}^{\nu }-
\frac{1}{2}\left(-g^{\mu \nu} \right)_Tk_{\perp}^2.
$$
In (\ref{eq:4}) $T_0(\zeta,W^2)$ is the amplitude corresponding to
the scattering of two photons with equal helicities,
$T_1(\zeta,W^2)$ denotes the amplitude with $L_z=\pm 1$, whereas
$T_2(\zeta,W^2)$ arises from the subprocess with opposite helicity
photons. In fact, in the collinear approximation, conservation of
the angular momentum along the collision axis leads to the
helicity conservation $h^{*}-h=h_1+h_2$, where $h^{*},\,h$ are the
helicities of the virtual and real photons and $h_1,\,h_2$ denote
the helicities of the produced quarks or gluons. When photons
produce a quark-antiquark pair, at the leading twist-2 level only
the subprocess with $L_z=0$ contributes to the amplitude. The
subprocess with $L_z=\pm 1$ is twist-3 and determines the
contribution $T_1(\zeta,W^2)$ appearing due to the interaction of
a longitudinally polarized virtual photon with the real one
\cite{Kiv01}. In the case where photons create a gluon pair,  both
the subprocesses with $L_z=0$ and $L_z=\pm 2$ contribute at the
twist-2 level to (\ref{eq:4}).

\begin{figure}
\epsfxsize=15 cm \epsfysize=5 cm \centerline{\epsffile{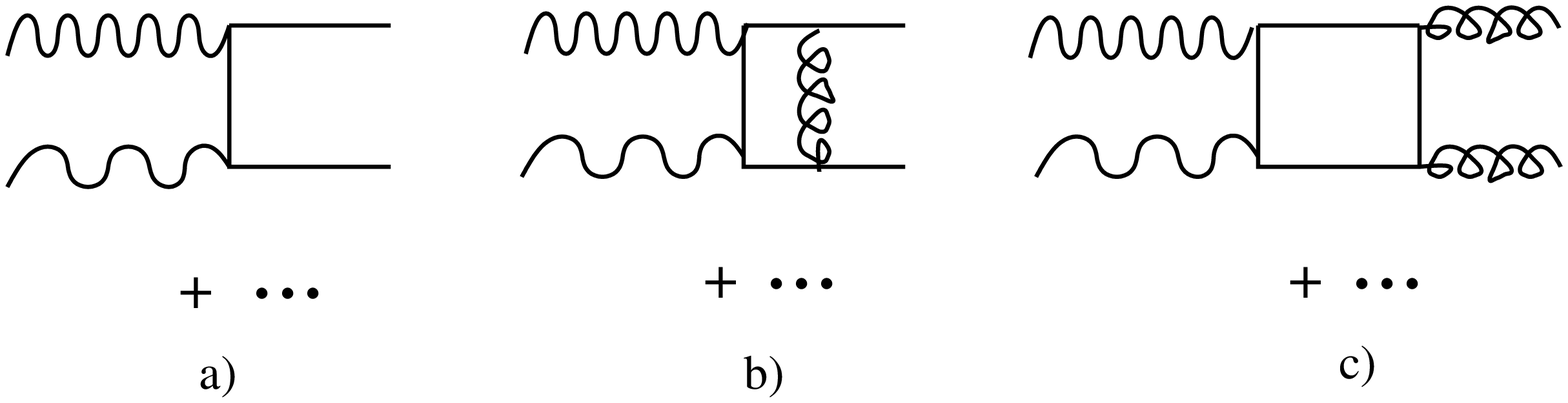}}
\caption{\label{fig:diag} Sample Feynman diagrams of the
hard-scattering subprocesses $\gamma^{*}\gamma \to q \overline{q}$
(at leading order- {\bf a}, at next-to-leading order -{\bf b}) and
$\gamma^{*}\gamma \to gg$ ({\bf c}).}
\end{figure}

We now focus on the amplitudes $T_i(\zeta,W^2),\, i=0, 2$ which do not
vanish at leading twist. They can be written as convolution integrals of
hard-scattering coefficient functions $C(z,\mu_F^2)$ and two-pion GDA's
$\Phi(z,\zeta,W^2,\mu_F^2)$, i.e

$$
T_0(\zeta,W^2)=\sum e_q^2
\int_0^1dzC_q(z,\mu_F^2)\Phi_q(z,\zeta,W^2,\mu_F^2)-
$$

\begin{equation}
\label{eq:5}
\sum e_q^2 \int_0^1dzC_g(z,\mu_F^2)\Phi_g(z,\zeta,W^2,\mu_F^2)
\end{equation}
and

\begin{equation}
\label{eq:6}
T_2(\zeta,W^2)=\int_0^1dzC_g^T(z,\mu_F^2)\Phi_g^T(z,\zeta,W^2,\mu_F^2),
\end{equation}
where $\mu_F^2$ is the factorization scale at which the hard and
soft parts of the reaction are defined. Here the coefficient
function $C_q(z,\mu_F^2)$ is calculated with NLO accuracy using
the subprocess $\gamma^{*} \gamma \to q \overline{q}$
(Figs.\ref{fig:diag}a,\ b) with $L_z=0$,  whereas the functions
$C_g(z,\mu_F^2)$ and $C_g^T(z,\mu_F^2)$ correspond to the
subprocess $\gamma^{*} \gamma \to gg$ (Fig. \ref{fig:diag}c) with
$L_z=0$ and $L_z=\pm 2$, respectively, and contribute only at the
next-to-leading order of PQCD due to quark-box diagrams.

The amplitudes $T_i(\zeta,W^2)$ do not depend on the
renormalization and factorization schemes and scales employed for
their calculation. But at any finite order of QCD perturbation
theory, due to truncation of the corresponding perturbation
series, the coefficient functions depend on both the factorization
$\mu _F^2$ and renormalization $\mu _R^2$ scales. An optimal
choice for these scales is always required to minimize higher
order corrections.
 The factorization scale
$\mu _F^2$ in exclusive processes is traditionally set equal to the hard momentum
transfer $Q^2$, and we shall follow this  prescription.

The functions $C(z)$  with NLO accuracy are given by the following
expressions \cite{Pol99}

\begin{equation}
 C_q(z)=C_q^0(z)+\frac{\alpha_{\rm
 s}(\mu_R^2)}{4\pi}C_q^1(z),
\label{eq:7}
\end{equation}

\begin{equation}
\label{eq:8}
C_g(z)=\frac{\alpha_{\rm s}(\mu_R^2)}{4\pi}C_g^1(z),\;\;
C_g^T(z)=\frac{\alpha_{\rm s}(\mu_R^2)}{4\pi}C_T^1(z),
\end{equation}
and

$$
C_q^0(z)=\frac{1}{\overline{z}}-\frac{1}{z},
$$

$$
C_q^1(z)=C_F\left [ \frac{\ln^2 \overline{z}}{z \overline{z}}-\frac{\ln^2 z}{z
\overline{z}}+\frac{\ln^2 z}{\overline{z}}-
\frac{\ln^2\overline{z}}{z} + 3\frac{\ln z}{\overline{z}}-
3\frac{\ln \overline{z}}{z}+\frac{9}{z}-\frac{9}{\overline{z}} \right ],
$$

$$
C_g^1(z)=\frac{1}{z^2 \overline{z}^2} \left [
z^2\ln^2z+\overline{z}^2\ln^2 \overline{z}+2z\overline{z}\ln
z\overline{z}-4z\ln z-4\overline{z}\ln \overline{z} \right ],
$$

\begin{equation}
\label{eq:9}
C_T^1(z)=\frac{2}{z\overline{z}},
\end{equation}
with $C_F=4/3$ being the color factor.

Choosing the renormalization scale $\mu_R^2$ is more subtle and we follow
the prescription \cite{BL83} that it is equal to the square of the momentum
transfer carrying by a virtual parton in each leading order Feynman diagram
of the underlying hard-scattering subprocess. Here, these scales are determined
by the leading order diagrams of the subprocess
$\gamma^{*} \gamma\to q\overline{q}$ and are given by the virtualities of the
off-shell fermion, which are equal to $Q^2z$ or to $Q^2 \overline{z}$
depending on the diagram.
In the present paper we adopt the symmetrized RC method, where
$\alpha_{\rm s}(Q^2z)$ and $\alpha_{\rm s}(Q^2\overline{z})$ are replaced by

\begin{equation}
\label{eq:11}
\alpha_{\rm s}(Q^2z),\;\alpha_{\rm s}(Q^2\overline{z}) \Rightarrow
\frac{\alpha_{\rm s}(Q^2z)+\alpha_{\rm s}(Q^2\overline{z})}{2}.
\end{equation}
The reasons which led to introduction of this version of the RC
method and further details have been presented in \cite{ag2,ag6}.

The next component in the  factorization formulas (\ref{eq:5}),
(\ref{eq:6}) is the generalized distribution amplitudes of the
$2\pi$ system. At present, little is known about these GDA's, but
using constraints originating from crossing symmetry and soft pion
theorems, as well as the evolution equation for GDA's, we can
model them. Further information on their analytic form should be
extracted from analysis of experimental data and corresponding
theoretical predictions and/or obtained, as in the case of the
usual DA's of mesons, via QCD non-perturbative methods.

The simplest $2\pi$ GDA's obtained  using the requirements
described above are

$$
\Phi_q(z, \zeta,
W^2,\mu_F^2)=20z\overline{z}(z-\overline{z})\frac{1}{n_f}M_q(\mu_F^2)A(\zeta,
W^2),
$$

\begin{equation}
\label{eq:12}
\Phi_g(z, \zeta,
W^2, \mu_F^2)=10z^2\overline{z}^2M_g(\mu_F^2)A(\zeta, W^2),
\end{equation}
where the model-dependent \cite{HPST} function $A(\zeta,W^2)$ is $z$
and $Q^2$ independent. Our results will not depend on the choice of this function.

The GDA's represented by the formula (\ref{eq:12})
depend on the momentum fractions carried by quarks $M_q(\mu_F^2)$ and
gluons $M_g(\mu_F^2)$ in the pion

$$
M_q(\mu_F^2)=M_q^{asy} \{1+R(\mu_0^2)L(\mu_F^2)
\},\;\;R(\mu_0^2)=\frac{M_q(\mu_0^2)-M_q^{asy}}{M_q^{asy}},
$$
with
\begin{equation}
\label{eq:13}
L(\mu_F^2)=\left [ \frac{\alpha_{\rm s}(\mu_F^2)}{\alpha_{\rm
s}(\mu_0^2)}\right
]^{\frac{\gamma_1^{+}}{\beta_0}},\;\;\; \gamma_1^{+}=\frac{2}{3}(n_f+4C_F),
\;\;\; M_g(\mu_F^2)=1-M_q(\mu_F^2),
\end{equation}
the asymptotic values of which are determined by the expressions

\begin{equation}
\label{eq:14}
M_q^{asy}=\frac{n_f}{n_f+4C_F},\;\;\; M_g^{asy}=\frac{4C_F}{n_f+4C_F}.
\end{equation}
For the helicity-two GDA $\Phi_g^{T}(z,\zeta,W^2, \mu_F^2)$ we take

\begin{equation}
\label{eq:15}
\Phi_g^{T}(z,\zeta,W^2, \mu_F^2)=D(\mu_F^2)z^2\overline{z}^2 A_g^{T}(\zeta, W^2),
\end{equation}
with

\begin{equation}
\label{eq:16}
D(\mu_F^2)=30D_{g}^{T}(\mu_0^2)\left[\frac{\alpha_{\rm s}(\mu_F^2)}{\alpha_{\rm
s}(\mu_0^2)}  \right
]^{\frac{\gamma^{TG}}{\beta_0}},\;\;\;\gamma^{TG}=7+\frac{2}{3}n_f.
\end{equation}
In (\ref{eq:13}), (\ref{eq:16}) $\mu_0^2$ and $D_g^T(\mu_0^2)$ are
the normalization point and constant, respectively.

\section{Borel resummed amplitudes }

Computation of the amplitudes $T_i(Q^2,\zeta,W^2)$ implies
integrations over $z$. Having inserted the explicit expressions of
the hard-scattering coefficient functions and the two-pion GDA's
into (\ref{eq:5}) and (\ref{eq:6}) we encounter divergences,
arising from the singularities of the coupling constant $\alpha
_{{\rm s}}(Q^2z)$ and $\alpha _{{\rm s}}(Q^2 \overline{z})$ in the
limits $z\rightarrow 0,\,1$. The  RC method  proposes a way to
cure these divergences.

To this end we express the running coupling
$\alpha _{{\rm s}}(Q^2z)$
in terms of $\alpha _{{\rm s}}(Q^2)$ [a similar argument holds also for
$\alpha_{\rm s}(Q^2\overline{z})$]. This
is achieved by applying the renormalization-group equation to
$\alpha _{{\rm s}}(Q^2z)$ \cite{st}. We get

\begin{equation}
\label{eq:17}
\alpha_{\rm s}(Q^2z)\simeq \frac{\alpha_{\rm s}(Q^2)}{1+\frac{\ln z}{t}}
\left[1-\frac{\alpha_{\rm s}(Q^2)\beta_1}{2\pi\beta_0}
\frac{\ln[1+\frac{\ln z}{t}]}{1+\frac{\ln z}{t}} \right].
\end{equation}
Here $\alpha_{\rm s}(Q^2)$ is the one-loop QCD coupling constant,
$t=4\pi / \beta_0 \alpha_{\rm s}(Q^2)=\ln \left( Q^2/\Lambda^2\right)$ and
$\beta_0,\,\beta_1$ are the QCD beta function one- and two-loop coefficients,
respectively,

$$
\beta_0=11-\frac{2}{3}n_f,\;\;\beta_1=51-\frac{19}{3}n_f,
$$
and $n_f$ is the number of active quark flavors. Equation
(\ref{eq:17}) expresses $\alpha_{\rm s}(Q^2z)$ in terms of
$\alpha_{\rm s}(Q^2)$ with an $ \alpha_{\rm s}^2(Q^2)$ order
accuracy.

Inserting (\ref{eq:17}) into the amplitudes
 we obtain integrals, which  can be
regularized and calculated using the method described in
\cite{ag4}. The amplitudes $T_i(Q^2,\zeta,W^2)$ are then written
as  perturbative series in $\alpha_{\rm s}(Q^2)$ with factorially
growing coefficients $C_n \sim (n-1)!$. Their resummation is
performed by employing the Borel integral technique \cite{TZ}.
Namely, one has to determine the Borel transforms
$B[T_i(Q^2,\zeta,W^2)](u)$ of the corresponding series and in
order to find the resummed expression for the amplitudes, has to
invert $B[T_i(Q^2,\zeta,W^2)](u)$ to get

$$
\left[T_i(Q^2,\zeta,W^2) \right]^{res} \sim
{\rm P.V.} \int_0^{\infty}du \exp \left[-\frac{4\pi u}{\beta_0\alpha_{\rm s}(Q^2)}\right]
B\left[T_i{(Q^2,\zeta,W^2)}\right](u)
$$

\begin{equation}
\label{eq:18}
+\left[T_i(Q^2,\zeta,W^2) \right]^{amb}
\end{equation}

The Borel transforms $B\left[T_i(Q^2,\zeta,W^2)\right](u)$ contain
poles $\{ u_0 \}$ located at the positive $u$ axis of the Borel
plane, which are exactly the IR renormalon poles. Therefore, the
inverse Borel transformation can be computed after regularization
of these pole singularities, which is achieved through a principal
value prescription. But the principal value prescription itself
generates higher twist (HT) ambiguities (uncertainties), which in
the right-hand side of (\ref{eq:18}) are denoted by
$[T_i(Q^2,\zeta,W^2)]^{amb}$. They are determined by the residues
of the Borel transforms at the IR renormalon poles $q \in \{u_0
\}$ and depend also on unknown numerical coefficients $\{N_q \}$

\begin{equation}
\label{eq:18a}
\left[T_i(Q^2,\zeta,W^2) \right]^{amb} \sim \sum_{q \in \{u_0 \}}N_q
\frac{\Phi_i^q(Q^2,\zeta,W^2)}{Q^{2q}}.
\end{equation}
The ambiguity (\ref{eq:18a}) can be used to estimate higher twist
corrections to the amplitudes stemming from other sources [for example, from
the $2 \pi$ higher twist GDA's].

A useful way to avoid the intermediate operations and obtain
directly the Borel resummed expressions is to introduce the
inverse Laplace transformations of the functions in (\ref{eq:17}),
i.e.

\begin{equation}
\label{eq:19}
\frac{1}{(t+x)^\nu}=\frac{1}{\Gamma(\nu)}\int_0^{\infty}du
\exp[-u(t+x)]u^{\nu-1}, \;\; Re\nu>0,
\end{equation}
and

\begin{equation}
\label{eq:20}
\frac{\ln [t+x]}{(t+x)^2}=\int_0^{\infty}du\exp[-u(t+x)](1-\gamma_E- \ln u)u,
\end{equation}
where $\Gamma(\nu)$ is the Gamma function, $\gamma_E \simeq 0.577216$ is the
Euler constant and $x=\ln z$ [$x=\ln \overline{z}$ in the case
$\alpha_{\rm s}(Q^2 \overline{z})$]. Then,  the QCD coupling
$\alpha_{\rm s}(Q^2z)$ may be written as \cite{ag3}

\begin{equation}
\label{eq:21}\alpha _{{\rm s}}(Q^2z)=\frac{4\pi }{\beta _0}%
\int_0^\infty due^{-ut}R(u,t)z^{-u},
\end{equation}
with

\begin{equation}
\label{eq:22}R(u,t)=1-\frac{2\beta _1}{\beta _0^2}u(1-\gamma _E-\ln t-\ln
u).
\end{equation}
The expression for the QCD running coupling (\ref{eq:21}) is
obtained from (\ref{eq:17}) and is suited to account for the
end-point effects. It differs from that introduced to perform the
resummation of diagrams with quark vacuum insertions (``bubble
chains'') into a gluon line \cite{ben,gr}. In exclusive processes
both these sources lead to power corrections. As noted above, in
the present work we consider contributions to the process
amplitudes arising only from the end-point regions.

Using (\ref{eq:12}), (\ref{eq:15}),  (\ref{eq:21}) and performing
the integration over $z$ we get\footnote {Below, for brevity, we
do not write down explicitly the higher twist ambiguities in the
resummed expressions.}

$$
[T_{0}(Q^2,\zeta,W^2)]^{res}=20A(\zeta,W^2)\sum e_q^2\left\{ \frac{M_q(Q^2)}{3n_f} \left
[1+ \frac {3C_F}{\beta_0}\int_0^\infty due^{-ut}R(u,t)Q(u) \right ] - \right.
\nonumber
$$
\begin{equation}
~~~~~~~~~~~~~~~~~~~ \left. \frac{M_g(Q^2)}{2\beta_0}\int_0^{\infty}due^{-ut}R(u,t)G(u) \right \},
\label{eq:23}
\end{equation}
and
\begin{equation}
\label{eq:24}
[T_2(Q^2,\zeta,W^2)]^{res}=2A_g^{T}(\zeta, W^2)\frac{D(Q^2)}
{\beta_0}\int_0^{\infty}due^{-ut}R(u,t)B(2-u,2),
\end{equation}
with

\begin{eqnarray}
\label{eq:25}
Q(u)= &\frac{\partial ^2}{\partial \beta ^2}B(2-u,\beta)|_{1}+
\frac{d^2}{d\beta ^2}B(2,\beta )|_{1-u}- \frac{\partial ^2}{\partial \beta
^2}B(1-u,\beta)|_2-
\frac {d^2}{d\beta ^2}B(1,\beta )|_{2-u}\ + \nonumber \\
&\frac{\partial ^2}{\partial \beta ^2}B(1-u,\beta)|_3+\frac{d^2}{d\beta
^2}B(1,\beta)|_{3-u}-\frac{\partial ^2}{\partial \beta ^2}B(2-u,\beta)|_{2}-
\frac{d^2}{d\beta ^2}B(2,\beta)|_{2-u}\ + \nonumber \\
&3\frac{\partial }{\partial \beta }B(1-u,\beta)|_3+3\frac{d}{d\beta
}B(1,\beta)|_{3-u}-3\frac{\partial }{\partial \beta }B(2-u,\beta)|_{2}-
3\frac{d}{d\beta }B(2,\beta)|_{2-u}\ -  \nonumber \\
&9B(3-u,1) -9B(1-u,3)+18B(2-u,2),
\end{eqnarray}
and
\begin{eqnarray}
G(u)=&\frac{\partial ^2}{\partial \beta ^2}B(1-u,\beta )|_{3}+\frac{d^2} {d
\beta ^2}B(1,\beta )|_{3-u}+2 \frac{\partial}{\partial \beta}B(2-u,\beta)|_2\ +\nonumber\\
 &2\frac{d}{d \beta}B(2,\beta)|_{2-u}-4\frac{\partial}{\partial
 \beta}B(1-u,\beta)|_2-4\frac{d}{d\beta}B(1,\beta)|_{2-u},
\label{eq:26}
\end{eqnarray}
where $B(x,y)$ is the Beta function $B(x,y)=\Gamma (x)\Gamma (y)/\Gamma(x+y)$.

In order to proceed one has to reveal the IR renormalon poles in
the resummed expressions. The analysis of  the pole structure of
$Q(u)$ and $G(u)$ is straightforward. The result is that the
function $Q(u)$ contains a finite number of triple poles located
at $u_0=1,\,2,\,3$, an infinite number of double poles at the
points $u_0=2,\,3,\,4\, \ldots \infty$ and single ones at the
points $u_0=1,\,2,\,3,\,4\,\ldots \infty$. For the function $G(u)$
we get: triple pole with location at $u_0=3$, infinite number of
double ($u_0=2,\,3,\,4\,\ldots \infty)$ and single poles
($u_0=1,\,2,\,3,\,4\,\ldots \infty)$. The  amplitude
$[T_2(Q^2,\zeta,W^2)]^{res}$ possesses only single poles at
$u_0=2,\,3$. In other words, by  employing (\ref{eq:17}) we have
transformed the end-point divergences in (\ref{eq:5}) and
(\ref{eq:6}) into the IR renormalon pole divergences in
(\ref{eq:23}) and (\ref{eq:24}). One can see that the integrals in
these expressions are the inverse Borel transformations
(\ref{eq:18}), where the Borel transforms
$B_{q(g)}[T_{0}(Q^2,\zeta,W^2)](u)$ of the quark and gluon
components of the amplitude $[T_0(Q^2,\zeta,W^2)]^{res}$ (in the
quark case the NLO part) and that of the amplitude
$[T_2(Q^2,\zeta,W^2)]^{res}$ up to constant factors are defined as

$$
B_{q(g)}[T_{0}(Q^2,\zeta,W^2)](u)\sim R(u,t)Q(u)\left[-G(u)\right],\; \;
B[T_2(Q^2,\zeta,W^2)](u)
\sim R(u,t)B(2-u,2).
$$

After removing IR renormalon divergences from (\ref{eq:23}) and
(\ref{eq:24}) by means of the  principal value prescription, they
determine the resummed amplitudes $[T_i(Q^2,\zeta,W^2)]^{res}$.
The final expressions $[T_i(Q^2,\zeta,W^2)]^{res}$ contain
power-suppressed corrections $\sim 1/Q^{2n},\,\,n=1,2,3 \ldots$ to
the amplitudes \cite{ag3,ag2,ag6} and are the main results of the
present work.

Let us now check the asymptotic limit of the resummed amplitudes.
In the asymptotic limit $Q^2\rightarrow \infty $ , GDA's
$\Phi_{q}(z,\zeta,W^2,Q^2)$ and $\Phi_{g}(z,\zeta,W^2,Q^2)$ evolve
to their asymptotic forms obtainable from (\ref{eq:12}) by means
of the replacements $M_q(Q^2) \to M_q^{asy}$ and $M_g(Q^2) \to
M_g^{asy}$. We need also to take into account that in this limit
the subleading term in the expansion of $\alpha _{{\rm s}}(Q^2z)$
through $\alpha _{{\rm s}}(Q^2)$ should be neglected, {\it i.e.}
\begin{equation}
\label{eq:27}\int_0^\infty due^{-ut}R(u,t) \to
\int_0^\infty due^{-ut}.
\end{equation}
Then the amplitude $[T_0(Q^2,\zeta,W^2)]^{res}$ takes the following form

\begin{equation}
\label{eq:28}
[T_{0}(Q^2,\zeta,W^2)]^{res}=\frac{20A(\zeta, W^2)}{3(n_f+4C_F)}\sum e_q^2\left\{ 1+ \frac {3C_F}{\beta
_0}\int_0^\infty due^{-ut}\left [Q(u)-2G(u) \right ]\right\}.
\end{equation}
The asymptotic limit of the integrals can be computed using
techniques, described in a detailed form in \cite{ag2,ag6}. After
some manipulations, one gets for the asymptotic limit of the
amplitude $[T_{0}(Q^2,\zeta,W^2)]^{res}$

\begin{equation}
\label{29}
\left [T_0(Q^2,\zeta,W^2) \right ]^{res} \to  \frac{20A(\zeta, W^2)}{3(n_f+4C_F)} \sum e_q^2
\left \{ 1-\frac{87}{9}C_F\frac{\alpha_{\rm s}(Q^2)}{4 \pi}\right \}.
\end{equation}

This expression coincides with the corresponding result from
\cite{Pol99}\footnote{Our definition of the function
$A(\zeta,W^2)$ differs by a factor $-1/6$ from that of
\cite{Pol99}.} and can be readily obtained within the standard
approach employing the $2\pi$ asymptotic GDA's. The  asymptotic
limit of the amplitude $[T_2(Q^2,\zeta,W^2)]^{res}$, due to the
factor $D(Q^2)$, is equal to zero.

This analysis of the asymptotic limit of the Borel resummed
amplitudes shows the internal consistency of the RC method itself.

\section{Numerical results}
Let us now present numerical estimates of the power corrections to
the amplitudes.
The resummed amplitude $[T_0(Q^2,\zeta, W^2)]^{res}$, which contains both the
hard perturbative component and power corrections, can be rewritten in the
form

\begin{equation}
\label{eq:30}
[T_0(Q^2,\zeta, W^2)]^{res}=T_0^{LO}(Q^2,\zeta, W^2)+T_0^{NLO}(Q^2,\zeta,
W^2)+T_0^{PC}(Q^2,\zeta, W^2),
\end{equation}
where the first two terms in the RHS of (\ref{eq:30}) are the LO
and NLO parts of the amplitude, whereas $T_0^{PC}(Q^2,\zeta, W^2)$
denotes the power corrections to it. The latter is given by the
expression

\begin{equation}
\label{eq:31}
T_0^{PC}(Q^2,\zeta, W^2)= \left [T_0(Q^2,\zeta, W^2) \right ]_{NLO}^{res}-
T_0^{NLO}(Q^2,\zeta, W^2).
\end{equation}
For our purposes it is convenient to normalize the expressions
(\ref{eq:30}) and (\ref{eq:31}) in terms of $T_0^{LO}(Q^2,\zeta,
W^2)$, which results in ratios independent on the function
$A(\zeta, W^2)$:

\begin{figure}
\epsfxsize=10 cm
\epsfysize=8 cm
\centerline{\epsffile{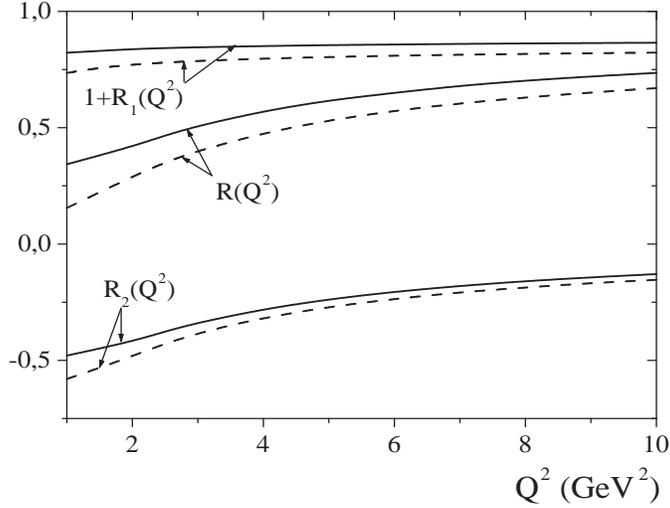}}
\caption{\label{fig:fig3}  The ratios $R_2(Q^2)$, $R(Q^2)$ and $1+R_1(Q^2)$ as
functions of $Q^2$. The solid (dashed) curve corresponds to the parameter
$M_q(1\ {\rm GeV}^2)=0.6$ [$M_q(1\ {\rm GeV}^2)=0.5$].}
\end{figure}

\begin{equation}
\label{eq:32}
R(Q^2)=1+R_1(Q^2)+R_2(Q^2).
\end{equation}
Here

\begin{equation}
\label{eq:33}
R_1(Q^2)=\frac{T_0^{NLO}(Q^2,\zeta, W^2)}{T_0^{LO}(Q^2,\zeta, W^2)},\;\;
R_2(Q^2)=\frac{T_0^{PC}(Q^2,\zeta, W^2)}{T_0^{LO}(Q^2,\zeta, W^2)}.
\end{equation}
In our calculations we
use the following values of the parameters $\Lambda$ and $\mu_0$

\begin{equation}
\label{eq:34}
\Lambda_4=0.2\ {\rm GeV}, \;\;\; \mu_0^2=1\ {\rm GeV}^2.
\end{equation}
To clarify the sensitivity of the predictions to the parameter
$M_q(\mu_0^2)$ we shall take the two plausible values
$M_q(1\ {\rm GeV}^2) = 0.5$ and $M_q(1\ {\rm GeV}^2) = 0.6$.
We use the two-loop approximation
for the QCD coupling $\alpha_{\rm s}(Q^2)$.

The amplitude $[T_0(Q^2,\zeta, W^2)]^{res}$
contains an infinite number of  IR renormalon poles. We  truncate
the corresponding series at some $n_{max}=50$ which is amply sufficient \cite{ag2,ag6}.

In Fig.\ \ref{fig:fig3} we show $R_2(Q^2)$ as a function of $Q^2$. The power corrections  amount to
some 50-60 per cent of the corresponding leading order
contribution at $Q^2=1\ {\rm GeV}^2$. They are not completely negligible also at $Q^2=10\ {\rm GeV}^2$
reaching around $15\%$  of the LO term. One  observes
that in the region $1\ {\rm GeV}^2 \leq Q^2 \leq 4\ {\rm GeV}^2$ the function
$R_2(Q^2)$ is more sensitive to the chosen value of the parameter $M_q$ than
in the domain $Q^2 \sim 10\ {\rm GeV}^2$.

In Fig.\ \ref{fig:fig3} the ratios $R(Q^2)$, $1+R_1(Q^2)$
are also shown. As is seen the power corrections significantly reduce the
amplitude $T_0(Q^2,\zeta, W^2)$ and this effect depends on the $2\pi$ GDA used
in calculations. Thus, at $Q^2=1\ {\rm GeV}^2$ the resummed amplitude
computed using the $2\pi$ GDA with the input parameter $M_q=0.6$ is
approximately twice as large as the same amplitude found employing the
GDA with $M_q=0.5$. At the higher values of $Q^2$ this difference
becomes more moderate $\sim 1.1$ at $Q^2=10\ {\rm GeV}^2$.

Another conclusion, which can be made  after analysis of Fig.\ \ref{fig:fig3}
is that the difference between the resummed [the ratio $R(Q^2)$] and the standard
 predictions for the amplitude [the ratio $1+R_1(Q^2)$]
becomes smaller at higher values of the momentum transfer $Q^2$. In fact, at
$Q^2=1\ {\rm GeV}^2$ the resummed amplitude is equal to $0.41$ of the standard
 result, whereas at $10\ {\rm GeV}^2$ one gets $0.85$
[for $M_q(1\ {\rm GeV}^2)=0.6$].

\begin{figure}
\epsfxsize=10 cm
\epsfysize=8 cm
\centerline{\epsffile{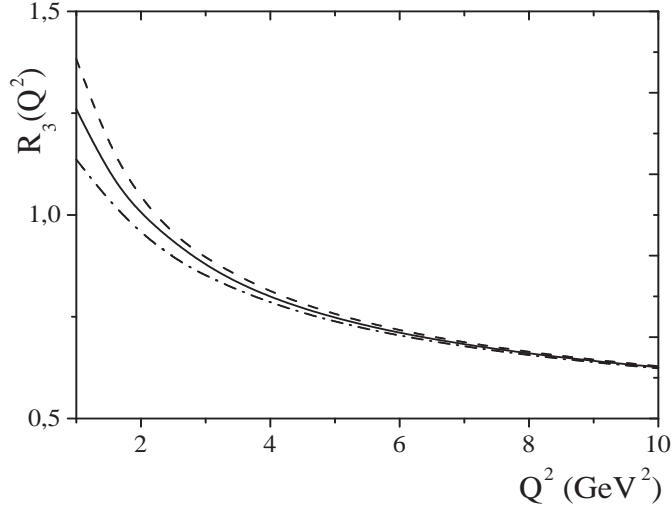}}
\caption{\label{fig:fig4}  The ratio $R_3(Q^2)$ vs $Q^2$. The solid line
corresponds to $R_3(Q^2)$ without the HT ambiguities. The broken lines are
obtained by taking into account the HT ambiguities (\ref{eq:18a}). For the
dashed line: $N_2=N_3=1$; for the dot-dashed line: $N_2=N_3=-1$.}
\end{figure}

\begin{figure}
\epsfxsize=10 cm
\epsfysize=8 cm
\centerline{\epsffile{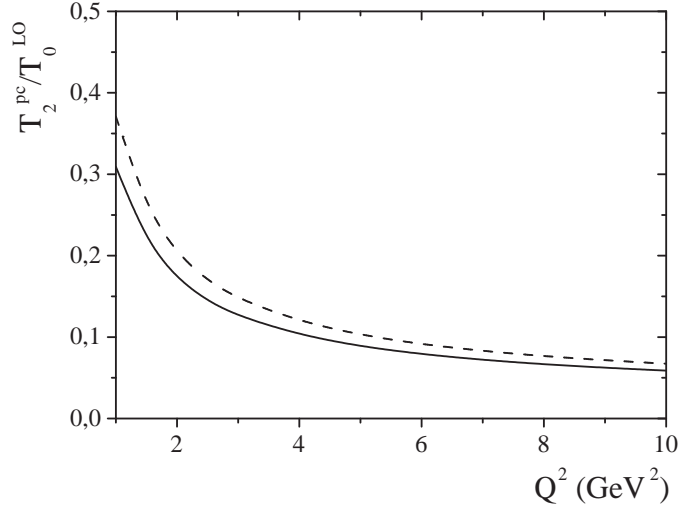}}
\caption{\label{fig:fig5}  The
ratio $T_2^{PC}/T_0^{LO}$ vs $Q^2$. The solid (dashed) curve
corresponds to the parameter
$M_q(1\ {\rm GeV}^2)=0.6$ [$M_q(1\ {\rm GeV}^2)=0.5$].}
\end{figure}

For phenomenological applications it is useful to parametrize the ratio
$R_2(Q^2)$ using the power-suppressed terms $\sim 1/Q^{2n},\ n=1,2,3$.
Our fitting procedure leads to the following expressions

$$
R_2(Q^2) \simeq \frac{1}{Q^2} \left [ -1.709+\frac{1.881}{Q^2}-\frac{0.7524}{Q^4}
\right ],\;\;\; M_q(1\ {\rm GeV}^2)=0.5,
$$

\begin{equation}
\label{eq:35}
R_2(Q^2) \simeq \frac{1}{Q^2} \left [-1.462+\frac{1.515}{Q^2}-\frac{0.533}{Q^4}
\right ],\;\;\; M_q(1\ {\rm GeV}^2)=0.6.
\end{equation}
The power corrections to the amplitude $T_2(Q^2,\zeta, W^2)$ are given by the formula

\begin{equation}
\label{eq:36}
T_2^{PC}(Q^2,\zeta, W^2)=\left [T_2(Q^2,\zeta, W^2)\right ]^{res}-T_2(Q^2,\zeta, W^2).
\end{equation}
 The ratio
$$
R_3(Q^2)=\frac{T_2^{PC}(Q^2,\zeta, W^2)}{T_2(Q^2,\zeta, W^2)}
$$
is shown in Fig.\ \ref{fig:fig4}. It turns out that in this
estimate of the power corrections to the amplitude
$T_2(Q^2,\zeta,W^2)$ are large and may still amount to a 60 per
cent increase of the amplitude at $Q^2 \sim 10\ GeV^2$. Such a
large magnitude of the end-point effects can be traced back to the
fact that $T_2(Q^2,\zeta,W^2)$ begins at $O(\alpha_{S}(Q^2))$. At
the same time, the ratio of these power corrections to the total
amplitude of the process remains within reasonable limits. To see
this, we normalize the corrections $T_2^{PC}$ in terms of the
$T_0^{LO}$ ignoring the different tensor factors in (\ref{eq:4})
and setting $A_g^{T}(\zeta,W^2)=A(\zeta,W^2)$, $D_g^T(1\ {\rm
GeV}^2)=1$, but keeping the factor $\sum e_q^2$ in definition of
the function $T_0(Q^2,\zeta,W^2)$. The ratio $T_2^{PC}/T_0^{LO}$
calculated in this approximate way is  shown in Fig.\
\ref{fig:fig5}. We find that power corrections $T_2^{PC}$ may
amount to $31-38\ \%$ of the $T_0^{LO}$ at $Q^2=1\ {\rm GeV}^2$
and only to $6-7\ \%$ of its value at $Q^2=10\ {\rm GeV}^2$. The
precise estimate of the effects generated by the helicity-two
component of the amplitude (\ref{eq:4}) requires more detailed
investigation.

The HT ambiguities (\ref{eq:18a}) coming from the principal value
prescription, in the process under consideration are sizeable only for small
values of the momentum transfer
$1\ {\rm GeV}^2 \leq Q^2 \leq 2.5\ {\rm GeV}^2$ [from $\pm 9 \%$ to $ \pm 3 \%$].
At $Q^2 =5\ {\rm GeV}^2$ they are already less than $\pm 1 \%$ of the
original result. As an example, the relevant curves for the ratio $R_3(Q^2)$
are shown in Fig.\ \ref{fig:fig4}.

\section{Concluding remarks}

In this work we have estimated the power corrections to the
amplitudes $T_i(Q^2,\zeta, W^2)$ of the process $\gamma^{*}\gamma
\to \pi \pi$, originating from the end-point regions $z\rightarrow
0,1$. To this end, we have employed the symmetrized RC method
combined with techniques of the IR renormalon calculus. We have
obtained the Borel resummed expressions $[T_i(Q^2,\zeta,
W^2)]^{res}$ for the amplitudes and  have removed IR renormalon
divergences by means of a principal value prescription. In the
considered process the Borel transform of the amplitude
$T_0(Q^2,\zeta, W^2)$ contains an infinite number of the IR
renormalon poles. Since  each IR renormalon pole $u_0=n$ in the
Borel transforms $B_{q(g)}[T_0(Q^2,\zeta, W^2)](u)$,
$B[T_2(Q^2,\zeta, W^2)](u)$ corresponds to the power correction
$\sim 1/Q^{2n}$ to the amplitudes, the expression (\ref{eq:23}),
in general, contains power corrections $\sim
1/Q^{2n},\,n=1,2,...\infty $. In numerical computations we have
truncated the corresponding series at $n_{\max }=50$. As an
important consistency check, we have proved that the result
obtained within the symmetrized RC method in the asymptotic limit
$Q^2\rightarrow \infty $ reproduces the standard  prediction for
the amplitudes.

It is known that the principal value prescription generates higher twist uncertainties.
We have shown that these uncertainties at $Q^2=1\ {\rm GeV}^2$ do not exceed $\pm 10 \%$
of the original prediction and can be safely neglected in the region
$Q^2 \geq 5\ {\rm GeV}^2$.

Our numerical calculations have demonstrated in an admittedly method
dependent way that the power corrections coming from the analysis of end-point regions
may be essential  in the region of photon virtualities
$Q^2 \sim {\rm a\ few}\ {\rm GeV}^2$. Therefore, the phenomenological analysis of the process
$\gamma^{*}\gamma \to \pi\pi$ in the presently experimentally-accessible
energy regimes should include them.

\bigskip

{\bf ACKNOWLEDGEMENTS }

\bigskip

We acknowledge useful discussions and correspondence with M. Diehl, G. Grunberg,
L. Szymanowski and O.V. Teryaev. One of the authors (S.~S.~A) would like
to thank the members of the PHASE group for hospitality at IPN Orsay
and gratefully to acknowledge the financial support from NATO Fellowship program.

\end{document}